\documentclass[12pt]{article} 

\usepackage{a4,amsmath,parskip,float}
\usepackage{epsfig}

\newcommand{\ppp}[1]{%
        \setbox0=\hbox{#1}%
        \kern-.02em\copy0\kern-\wd0
        \kern+.04em\copy0\kern-\wd0
        \kern-.02em\raise.0217em\box0}
\newcommand{\vek}[1]{
         \mathchoice{\mbox{\boldmath$#1$}}%
        {\mbox{\boldmath$#1$}}%
        {\ppp{$\scriptstyle#1$}}%
        {\ppp{$\scriptscriptstyle#1$}}}

\newcommand{\gr}[1]{\ensuremath{\Gamma_{\rho}({\vek b}-{\vek s}_{#1}})}

\newcommand{\ggr}[1]{\ensuremath{\Gamma_{\gamma^*\rho}({\vek b}-{\vek s}_{#1}})}

\newcommand{\intn}{\int \negthickspace}

\newcommand{\fr}[1]{\ensuremath{f_{\rho}(#1)}}
\newcommand{\frs}[1]{\ensuremath{f_{\rho}^{\ast}(#1)}}

\newcommand{\fgr}[1]{\ensuremath{f_{\gamma^*\rho}(#1)}}
\newcommand{\fgrs}[1]{\ensuremath{f_{\gamma^*\rho}^{\ast}(#1)}}

\newcommand{\Dr}{\ensuremath{\Delta_{\gamma^*\rho}\,}}

\newcommand{\lsim}{$\raisebox{-0.8ex} {$\stackrel{\textstyle <}{\sim}$}$}
\newcommand{\gsim}{$\raisebox{-0.8ex} {$\stackrel{\textstyle >}{\sim}$}$}

\newcommand{\T}[1]{{\mathrm{#1}}}

\begin{document}  
\begin{titlepage}
\vspace*{-2cm}
\begin{flushright}
\bf 
TUM/T39-00-07
\end{flushright}

\bigskip

\begin{center}
{\large\bf Coherence Effects  in Diffractive Electroproduction 
\protect \\ of $\rho$ Mesons  from Nuclei}

\vspace{2.cm}

{\large T.~Renk, G.~Piller and W. Weise $^{*}$}
\date{\today{}}

\vspace{2.cm}

Physik Department\\ Technische Universit\"{a}t M\"{u}nchen \\
D-85747 Garching, Germany 

\vspace*{3cm}

{\bf Abstract}
\bigskip

\begin{minipage}{15cm}
A systematic multiple scattering formalism for vector meson
electroproduction from nuclei is developed, with emphasis on the
formation, propagation and hadronization scales for quark-gluon
fluctuations of the virtual high-energy photon.
The theory is compared to HERMES measurements of $\rho$
electroproduction on $^{14}$N. The nuclear transparency as a
function of the vector meson propagation length is well
reproduced.
\end{minipage}
\end{center} 

%\vspace*{1.cm}

\vspace*{1.cm}

\noindent $^{*}$) Work supported in part by BMBF and DFG.

\end{titlepage}

\section{Introduction}

Electroproduction of vector mesons from nuclei 
is an excellent tool to investigate the formation and propagation of 
quark-antiquark pairs under well-controlled kinematical conditions. 
The observed nuclear coherence effects provide important information on 
properties of quark-gluon configurations  
present in the wave function of the interacting virtual photon, and 
on their hadronization into the finally observed vector meson.
Recent  data on $\rho$ production from various nuclei have 
become available from measurements of the HERMES collaboration at 
DESY \cite{Ackerstaff:1999wt}. 
Earlier measurements at  higher energies were  performed 
at FNAL \cite{Adams:1995bw} and  CERN \cite{Arneodo:1994id}.

Consider first the electroproduction of vector mesons from free nucleons. 
The four-momentum of the interacting virtual photon is denoted by 
$q^{\mu}=(\nu,\vek q)$. 
We use  $Q^2 = -q^2$, and $x=Q^2/2M\nu$ is the 
Bjorken scaling variable expressed in the lab frame, $M$ is the nucleon 
mass. 
At large photon energies and small momentum transfers to the target
nucleon, diffractive production mechanisms dominate:  
the vector meson 
emerges from the reaction leaving the nucleon intact.

According to the current understanding, the picture for this
process is believed to be as follows:
At $\nu \geq 3\,$GeV  and $x \leq 0.1$ one can decompose 
the amplitude for the diffractive  production of a vector meson $V$  
in terms of colorless hadronic or quark-gluon fluctuations of the 
(virtual) photon (see e.g. \cite{Gribov70,Spital76}):
\begin{equation}
f^{\gamma^*N\rightarrow VN} = 
\sum_{h}{\langle 0|\epsilon_{\gamma^*}\cdot J^{em}|h\rangle 
\over E_{h}-\nu} \, f^{hN\rightarrow V N}.
\label{eq:spectral}
\end{equation}
Here $\epsilon_{\gamma^*}$ is the polarization vector of the 
virtual photon, $J^{em}$ denotes the electromagnetic current, and  
$E_{h}$ stands for  the energy of the intermediate hadronic state $h$.

At photon energies $3 < \nu < 30$ GeV  
and $Q^2 \,\lsim \,1$ GeV$^2$ ,
contributions to the photoproduction amplitude $f^{\gamma^*N\rightarrow VN}$ 
from hadronic states with large invariant mass $m_h$   
are  suppressed by large energy denominators 
$E_{h} - \nu  \approx  ({m_{h}^2 + Q^2})/({2 \nu})$.  
This restriction to light intermediate hadronic states  
implies vector meson dominance (see e.g. \cite{BSYP}): 
in the lab frame the photon converts into a vector meson prior to the 
scattering from the target. 

On the other hand, at $Q^2\gg 1$ GeV$^2$ and $x \ll 0.1$ 
perturbative QCD is believed to be applicable \cite{BFGMS}:  in the lab frame 
the photon first converts to a quark-antiquark-gluon jet which then 
interacts with the nucleon. At high photon energies  the finally observed vector
meson is formed at a much later stage.

The transition from small to large $Q^2$ interpolates between 
non-perturbative hadron formation and perturbative quark-antiquark-gluon 
dynamics, a question of central importance in QCD. 
Nuclear targets are particularly helpful at this point because they serve 
as analyzers for the coherent interaction of the produced $q\bar q$-gluon 
system with several nucleons \cite{Brodsky:1988xz}. 
The average distance between two nucleons provides the ``femtometer scale'' 
which can be used to measure the relevant coherence lengths 
(for reviews and references see 
\cite{Nikolaev:1992si,Frankfurt:1994hf,Jain:1996dd}).

The production process is driven by the following characteristic 
scales. 
The propagation length (for reviews and references see e.g. \cite{BSYP,PWrep}), 
\begin{equation}\label{coh}
\lambda \approx \frac{2\nu}{m^2 + Q^2}\,,
\end{equation}
represents the longitudinal distance over which a hadronic fluctuation 
of invariant mass $m$ propagates in the lab frame when 
induced by a photon of energy $\nu$ and virtuality $Q^2$. 
At large $Q^2$ the initially produced wave packet is characterized 
by its transverse size $b$. For longitudinally 
polarized photons, detailed calculations suggest \cite{LNGS}
\begin{equation}\label{ej}
b  =  \frac{const}{Q}.
\end{equation}
In perturbative QCD the minimal ($q \bar q$) Fock space component has 
$const \sim 4$--$5$ at $Q^2 \,\gsim 
\linebreak
5 \,\T{GeV}^2$ 
\cite{Frankfurt:1996jw}. 
Thus for $Q^2 = 5\,\T{GeV}^2$  and $const = 4$, the 
transverse size of the initial wave packet is $b\simeq 0.35$ fm, 
much smaller than the diameter of a fully developed $\rho$ meson.

This paper is organized as follows. The multiple scattering theory for the 
vector meson electroproduction from nuclei, both coherent and incoherent, 
is developed in the next section. Section 3 presents results with detailed 
comparisons to the HERMES measurements, using $^{14}$N targets, and a
discussion of the observed transparency and propagation length effects. 
A concluding summary is given in Section 4.

\section{Amplitudes and cross sections} 
\label{sec:amplitudes_cross sections}

In the following we derive amplitudes and cross sections for 
exclusive electroproduction of vector mesons from nuclei. 
Although we concentrate on  $\rho$ mesons, 
our formalism can be extended  easily to other vector mesons.

At high energies the production of $\rho$ mesons is driven by diffraction: 
the incident photon interacts 
with one of the nucleons in the  nucleus and produces 
a hadronic state $h$, leaving the nucleon intact. 
The hadron  $h$ then propagates through the nuclear medium and 
possibly  re-scatters, again diffractively, from further nucleons, 
subsequently forming the observed $\rho$ meson.

%%%%%%%%%%%%%%%%%%%%%%%%%%%%%%%%%%%%%%%%%%%%%%%%%%%%%%%%%%%%%%%%%%%%
\begin{figure}[t]
\bigskip
\begin{center} 
\epsfig{file=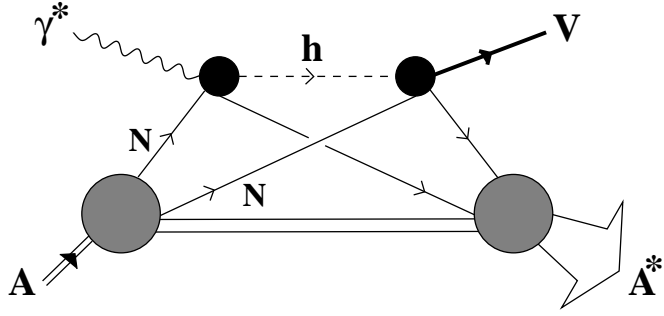,height=40mm}
\end{center}
\caption[...]{Double scattering contribution to (virtual) photoproduction of a 
vector meson.}
\label{double scatt}
\bigskip
\end{figure}
%%%%%%%%%%%%%%%%%%%%%%%%%%%%%%%%%%%%%%%%%%%%%%%%%%%%%%%%%%%%%%%%%%%%%%

The leading nuclear effects result from processes involving two nucleons 
as illustrated in Fig.\ref{double scatt}.  
The diffractive production of the intermediate hadronic state, 
$\gamma^* + N \rightarrow h + N$, is described by the amplitude 
$f_{\gamma^*h}$. At high energies this amplitude  depends, to a good approximation, only on
the momentum transfer $\vek k_t$ perpendicular to the direction of the 
incoming photon.
One introduces the profile function (see e.g. \cite{BSYP,Yennie71}):
\begin{equation}
\label{E-GammaToF}
\Gamma_{\gamma^* h}({\vek b}) = \frac{1}{2 \pi i |\vek q|}
\int d^2 k_t  \,f_{\gamma^* h}({\vek k}_t) \,
e^{i {\vek k}_t \cdot {\vek b}},   
\end{equation}
where the three-momentum $\vek q$ of the incident photon is chosen 
in the $z$-direction,   $\vek q = \hat{\vek z} |\vek q|$.
$\Gamma_{\gamma^* h}({\vek b})$ is a measure of the interaction strength at 
fixed impact parameter $\vek b$. 

In the double scattering process illustrated in Fig.~\ref{double scatt},  the
intermediate state $h$ interacts with 
a second nucleon, producing the final $\rho$ meson state, 
$h + N \rightarrow \rho + N$. 
The corresponding amplitude $f_{h\rho}$  defines a profile function 
$\Gamma_{h \rho}$ 
as in Eq.(\ref{E-GammaToF}).
In total, the double scattering contribution to $\rho$ production is 
described by the nuclear profile operator  
\begin{equation} \label{eq:gamma_ds}
{\Gamma}_{\gamma^*\rho}^{A (2)}(\vek b, \vek s_i, \vek s_j) = 
\sum_{\left\{i\ne j\right\}=1}^A   
\Gamma_{h\rho}(\vek b - \vek{s}_j) \, e^{i \Delta_{h \rho} z_j}
\,\theta(z_j-z_i)\,
\Gamma_{\gamma^* h}(\vek b - \vek{s}_i) \,e^{i \Delta_{\gamma^* h} z_i}.
\end{equation}
The two nucleons labeled $i$ and $j$  are located at impact parameters  
$\vek b - \vek s_{j,i}$ and at longitudinal distances 
$z_{i,j}$. The $\theta$-function makes sure 
that the intermediate hadronic state $h$ is produced before it 
re-scatters. 
The difference between the three-momenta of the incoming virtual photon 
and the hadronic state $h$ with invariant mass $m_h$  
leads to the phase $e^{i \Delta_{\gamma^* h} z_i}$. 
For large photon energies ($\nu^2 \gg Q^2$ as well as $\nu \gg m_{h})$  
transitions in the forward direction dominate,  
and one finds \cite{BSYP,Yennie71}:  
\begin{equation} \label{eq:Dgh}
\Delta_{\gamma^* h} \approx \frac{Q^2 + m_{h}^2}{2 \nu},  
\end{equation}
which corresponds to the inverse of the longitudinal propagation length
(\ref{coh}) of the intermediate system $h$. 
Similarly, the transition from the intermediate state $h$ to the final $\rho$ leads 
to the second phase in Eq.(\ref{eq:gamma_ds}), 
\begin{equation} \label{eq:Dhr}
\Delta_{h \rho} \approx \frac{m_{h}^2 - m_{\rho}^2}{2 \nu}.
\end{equation}
This phase carries the information about the time it takes to form the 
final vector meson out of the intermediate state $h$. In fact $\Delta_{h\rho}$ 
is just the inverse of that formation time, 
\begin{equation} \label{eq:ft}
\tau_f = \frac{1}{\Delta_{h\rho}} \simeq \frac{\nu}{m_{\rho} \delta m} 
\end{equation}
with $\delta m = m_h - m_{\rho}$.

The role played by the phases (\ref{eq:Dgh},\ref{eq:Dhr}) 
can be understood as follows: the nuclear scattering amplitude 
involves an  integration of the nuclear profile operator over the 
volume of the target nucleus. 
The production process is dominated by contributions 
with phases roughly constant over the nuclear 
volume, while terms with  rapidly oscillating phases are 
suppressed. 
This observation leads to the following scenarios
depending on $Q^2$:

\begin{itemize}

\item 
At large photon four-momenta ($Q^2 \gg 1$ GeV$^2$)  and small Bjorken-$x$
($x\ll 0.1$),   the longitudinal 
interaction length $\lambda =  \Delta_{\gamma^* h}^{-1}$ exceeds  
nuclear dimensions. 
Therefore, as seen from the lab frame, the photon first couples to a 
quark-antiquark pair which subsequently scatters from the nucleus. 
The $q\bar q$ pair has a chance to travel 
over large distances inside the nucleus and interacts only weakly, its cross 
section being proportional to $b^2 \sim 1/Q^2$. 
Since the formation time $\tau_f = \Delta_{h\rho}^{-1}$ is large 
in the considered kinematic region, 
the finally observed vector meson forms far outside the nucleus. 
This phenomenon is commonly referred to as color coherence or 
color (singlet) transparency 
\cite{Brodsky:1988xz,Nikolaev:1992si,Frankfurt:1994hf,Jain:1996dd}.

\item
At small photon virtualities, $Q^2 \,\lsim \,1$ GeV$^2$, 
vector meson production is driven by 
the interaction of hadronic fluctuations of the photon. 
In particular the contribution of the $\rho$ meson component in  
the photon spectral function dominates since it implies  
$\Delta_{h \rho} = 0$.
At high energies, $\nu \sim 30$ GeV, the $\rho$ meson propagation length 
exceeds the size of the nucleus. 
Therefore, the photon first couples to low mass   
($m_h \sim m_\rho$) hadronic states which then scatter from the nucleus. 
The interaction cross sections of these states 
are large, comparable to typical hadron-nucleon cross sections. 
As a consequence coherent multiple interactions with several nucleons 
lead to significant nuclear modifications of the 
production process.

\item
At $Q^2 \,\lsim \,1$ GeV$^2$ and moderate energies, e.g. 
$\nu \sim 4$ GeV, the longitudinal propagation length $\lambda$ 
for a $\rho$ meson is less than $1$ fm, so that  
the meson is produced on a nucleon inside the nucleus. 
Once produced, the meson  undergoes final state interactions 
with residual nucleons. 
Note however that the path of the vector meson through the nuclear medium is, 
on average, shorter than in the previous case, where hadronic states 
propagate through the nucleus as a whole. 
One therefore expects nuclear modifications of the $\rho$ meson production
cross section  at low photon energies to be less pronounced than 
at high energies.

\end{itemize}

Higher order multiple scattering contributions 
involve transition amplitudes $f_{h'h}$ between different hadronic states  
carrying photon quantum numbers. These, as well as the complete set of amplitudes 
$f_{h\rho}$ which enter already in double scattering, Eq.(\ref{eq:gamma_ds}), 
are in general unknown. 
Consequently, any calculation of vector meson production at high  
energies must involve some  model assumptions about $f_{h'h}$ 
(for examples see \cite{Nikolaev:1992si,Frankfurt:1994hf}). 
On the other hand, the recent HERMES data on exclusive $\rho$ production were
taken at moderate photon energies $\nu$, 
where the typical formation times $\tau_f$ are less than, or at most   
of the order of, the average nucleon-nucleon separation in nuclei. 
Contributions to the production process from 
intermediate $\rho$ mesons are therefore supposed to dominate. 
They lead to the nuclear profile operator \cite{BSYP,Yennie71}:
\begin{equation}
\label{E-Nprofile}
{\Gamma}_{\gamma^*\rho}^{A}(\vek b;\{\vek s\}) = 
\sum_{i=1}^A \bigg\{\prod_{j \neq i} \big[1-\gr{j} \theta(z_j-z_i)\big]
\bigg\} \,\ggr{i} \exp(i \Delta_{\gamma^*\rho} z_i),  
\end{equation}
which we use as a starting point for our investigations. 
Elastic $\rho$-nucleon scattering is described 
in Eq.(\ref{E-Nprofile}) by the profile function 
$\Gamma_{\rho N \rightarrow \rho N} \equiv \Gamma_\rho$. 
Its relation to the $\rho N$ scattering amplitude 
$f_{\rho N \rightarrow \rho N}\equiv f_\rho$ is
\begin{equation}
\Gamma_{\rho}({\vek b}) =
%\equiv \Gamma_{\rho\rho}({\vek b}) = 
\frac{1}{2 \pi i |\vek q|}
\int d^2 k_t  \,f_{\rho}({\vek k}_t) \,
e^{i {\vek k}_t \cdot {\vek b}}. 
\end{equation}

\subsection{Inclusive production}
\label{ssec:incl}

In the inclusive electroproduction  of $\rho$ mesons the nuclear 
final state is not observed. Summing over all possible nuclear 
excitations and applying closure leads to \cite{BSYP,Yennie71}
\begin{equation}
\label{E-InclusiveAnsatz}
\begin{split}
\frac{d \sigma_{\gamma^* A \rightarrow \rho X}}{d t}
=&
\frac{1}{4 \pi}
\intn d^2 b \intn d^2 b^{\prime}
%\exp[i {\vek k_t } \cdot({\vek b^{\prime}-\vek b})]
\,
e^{i {\vek k_t }\cdot({\vek b^{\prime}-\vek b})}
\,
\langle 0|{\Gamma}_{\gamma^*\rho}^{A\dagger}({\vek b^{\prime};\{\vek s\}})
{\Gamma}_{\gamma^*\rho}^{A}({\vek b;\{\vek s\}})|0\rangle.  
\end{split}
\end{equation}
Here $|0\rangle$ denotes the nuclear ground state;  
$t = (q - k)^2$ refers to the squared momentum transfer where  
$k^{\mu}$ is  the four-momentum of the produced $\rho$ meson. 

The next step is to expand both nuclear
profile operators  in Eq.(\ref{E-InclusiveAnsatz}) 
in terms of nucleon profile functions, using Eq.(\ref{E-Nprofile}). 
Collecting in ${\Gamma}_{\gamma^*\rho}^{A}$ and
${\Gamma}_{\gamma^*\rho}^{A\dagger}$ terms with products of at
most $n$ profile functions, $\Gamma_{\rho}$,  defines the cross sections 
\begin{equation} \label{eq:up_to_n}
\frac{d \sigma_{\gamma^* A \rightarrow \rho X}^{(n)}}{d t}.
\end{equation}
They describe processes  with up to  $2n+2$ nucleons involved in 
the production and re-scattering of the $\rho$-meson.
At the high energies involved, it is justified to neglect $NN$-correlations 
and to treat the nucleus as a system of $A$ independent nucleons, described by 
a density distribution $\rho_A(\vek r)$. The cross sections in
Eq.(\ref{eq:up_to_n}) 
are then expressed through nucleon amplitudes and nuclear form factors, 
\begin{equation} 
\label{eq:ff}
S_A(\vek k_t, \Delta) = \int d^2 b \,dz\,\rho_A(\vek b, z) \, 
e^{-i\left(\vek k_t \cdot \vek b + \Delta z\right)}. 
\end{equation}
In numerical calculations we use a   Gaussian parametrization 
for the nuclear one-body density, applicable to light nuclei:
\begin{equation}
\rho_A({\vek r}) = \left( \frac{3}{2 \pi \langle r^2\rangle_A} \right)^{3/2} 
\exp \left(- \frac{3 r^2}{2 \langle r^2\rangle_A} \right), 
\end{equation}
with normalization $\int d^3 r \rho_A(\vek r) = 1$ and 
mean square radius $\langle r^2\rangle_A = \int d^3 r r^2\rho_A(\vek r)$.

The so-called single scattering contribution describes 
the production of a  $\rho$ meson through the interaction of 
the virtual photon with one  nucleon inside the nucleus. 
Re-scattering from further nucleons does not occur here.
One finds: 
\begin{equation} \label{eq:incl_sing}
\frac{d \sigma_{\gamma^* A \rightarrow \rho X}^{(0)}}{d t}
= 
A\,\frac{d \sigma_{\gamma^* N \rightarrow \rho N}}{d t} 
\,\left[1 + (A-1) S_A({\vek k}_t, -\Dr)\,S_A(-{\vek k}_t, \Dr)  \right].  
\end{equation}
The first term represents the $A$ possibilities in which the nuclear profile
operator and its conjugate in Eq.(\ref{E-InclusiveAnsatz}) 
involve the same nucleon.
The second term counts $A (A-1)$ possibilities with $\Gamma^A$ and
$\Gamma^{A\dagger}$ involving different nucleons. 
This contribution is proportional to the square of the elastic nuclear form
factor (\ref{eq:ff}). 
As expected, the single scattering cross section 
reduces to $A^2$ times the exclusive $\rho$ production 
cross section from free nucleons in the limit of vanishing momentum transfer 
($\vek k_t, \Dr \rightarrow 0$).

The leading  correction to single scattering 
results when one of the profile operators in Eq.(\ref{E-InclusiveAnsatz}) 
involves the  re-scattering of the
$\rho$ meson from one nucleon in the target nucleus, while 
re-scattering terms are omitted in the second profile operator.
We find  
\begin{equation} \label{eq:incl_double}
\begin{split}
\frac{d \sigma_{\gamma^* A \rightarrow \rho X}^{(1)}}{d t} \,= \, &  
\frac{d \sigma_{\gamma^* A \rightarrow \rho X}^{(0)}}{d t}
-\frac{A(A-1)}{2 |\vek q|^3} \intn d^2 l
\quad \text{Im} 
\left[ \fgrs{{\vek k}_t} \fgr{{\vek k}_t - {\vek l}} \fr{{\vek l}} \right]
\\ & \times
\left\{
S_A({\vek l},0) S_A({-\vek l},0) + 
S_A({- \vek k}_t+{\vek l}, \Dr) S_A({\vek k}_t - {\vek l}, -\Dr)\right.
\\ 
& 
\;\;\;\;\;
\left.+
(A-2) S_A({\vek k}_t, -\Dr) S_A({- \vek k}_t + l, \Dr) S_A({-\vek l},0) 
\right\}
+ \dots ,
\end{split}
\end{equation}
where we have  suppressed  
terms resulting from the square of re-scattering amplitudes. 
They can be found in the Appendix and are included in the numerical 
calculations.

At high energies the nucleon amplitudes 
$f_{\gamma^* \rho}$ and $f_{\rho}$ are dominated by their 
imaginary parts.
Re-scattering therefore reduces the production cross section, an 
observation first made in \cite{Gribov70}. 
The first re-scattering term in Eq.(\ref{eq:incl_double}) 
is proportional to $S_A({\vek l},0) \,S_A({-\vek l},0)$ 
and does not depend on  the propagation length $\lambda = \Delta^{-1}_{\gamma^*\rho}$.
This describes $\rho$
production taking place on the same nucleon in both profile operators.
The remaining contributions correspond to the interference of amplitudes 
where the $\rho$ is produced on different nucleons. 
These terms are enhanced for large $\lambda$.

In our numerical calculations we have evaluated the inclusive 
production cross section (\ref{E-InclusiveAnsatz}) up to $n=2$. 
This includes terms which account for the re-scattering of the 
produced $\rho$ from up to $4$ nucleons. 
Explicit expressions are given in the Appendix.

\subsection{Coherent production}
\label{ssec:coh}

In coherent $\rho$ production the target nucleus stays intact, 
i.e. one considers the reaction  $\gamma^* + A \rightarrow \rho + A$, 
where the nucleus recoils as a whole and is left in its ground state. 
An experimental signature for this would be the detection of the
recoiling nucleus. In practice a situation like this is approximately
realized at momentum transfers smaller than the nuclear Fermi
momentum, $|t|<0.1$ GeV$^2$, where the nucleus
responds quasi-elastically to the incoming virtual photon. 

%The corresponding cross section is 
%%
%\begin{equation}
%\frac{d \sigma_{\gamma^* A \rightarrow \rho A}}{d t} =
%\frac{\pi}{|\vek q|^2} 
%\left|f_{\gamma^* A \rightarrow \rho A}({\vek k}_t)\right|^2.  
%\end{equation}
%
At high photon energies, the corresponding amplitude is related 
to the ground state matrix element of the nuclear 
profile function: 
\begin{equation}
\label{E-FToGamma}
f_{\gamma^* A \rightarrow \rho A} ({\vek k_t}) = 
\frac{i |\vek q| }{2 \pi}
\int d^2 b e^{- i {\vek k}_t\cdot {\vek b}} 
\langle 0|{\Gamma}_{\gamma^*\rho}^{A}({\vek b;\{\vek s\}})|0\rangle.
\end{equation}

As in the inclusive case, the $\rho$ is produced 
in a diffractive  interaction of the virtual photon with one of the 
nucleons in the nucleus, $\gamma^* + N \rightarrow \rho + N$. 
Then the $\rho$ meson propagates through the residual nuclear system 
where it may interact with other nucleons.
We expand the nuclear electroproduction 
amplitude (\ref{E-FToGamma}) according to the number of 
$\rho$-nucleon re-scattering processes involved, following Eq.(\ref{E-Nprofile}):
\begin{equation} \label{eq:coh_exp}
f_{\gamma^* A \rightarrow \rho A} = 
\sum_{n=0}^{A-1}
f_{\gamma^* A \rightarrow \rho A}^{(n)}.  
\end{equation}
The amplitude $f_{\gamma^* A \rightarrow \rho A}^{(n)}$ 
includes  all contributions to $f_{\gamma^* A \rightarrow \rho A}$ 
with products of $n$ nucleon profile functions $\Gamma_\rho$ or, 
equivalently, $\rho$-nucleon scattering amplitudes $f_{\rho}$. 
The  cross sections 
\begin{equation} \label{eq:ms_coh}
\frac{d \sigma_{\gamma^* A \rightarrow \rho A}^{(n)}}{d t} =
\frac{\pi}{|\vek q|^2} 
\left|\sum_{i=0}^{n}f_{\gamma^* A \rightarrow \rho A}^{(i)}({\vek k}_t)\right|^2.  
\end{equation}
describe the electroproduction of a $\rho$ meson and its subsequent 
re-scattering from up to $n$ nucleons.

The single scattering amplitude    
$f_{\gamma^* A \rightarrow \rho A}^{(0)}$ has  no re-scattering. 
Within the independent particle description for the target nucleus 
one obtains:
\begin{equation} \label{eq:coh_sing}
f_{\gamma^* A \rightarrow \rho A}^{(0)}({\vek k}_t) = 
A \,  S_A({\vek k}_t, -\Dr) \fgr{{\vek k}_t}.
\end{equation}
The nuclear form factor in  Eq.(\ref{eq:coh_sing})  accounts 
for the probability that the target nucleus 
stays intact during the interaction. 
$S_A$ drops sharply for momenta much larger than the inverse 
nuclear radius, $|\vek k_t| > 1/R_A$,  
so that  the single scattering contribution to 
coherent $\rho$ production is small for $|t| > 1/R_{A}^2$.

If the produced $\rho$ meson re-scatters from one nucleon after its 
production,  one gets: 
\begin{eqnarray} \label{eq:coh_ds}
f_{\gamma^* A \rightarrow \rho A}^{(1)}({\vek k_t}) &=
&\frac{A(A-1)i}{4 \pi |\vek q|}
\intn d^2 l 
\,
f_{\rho}\left({\vek k}_t/2 + \vek l\right) 
\,
f_{\gamma^*\rho}\left({\vek k}_t/2 - {\vek l}\right)
\nonumber \\
&&\hspace*{3cm}
\times
S_A\left({\vek k}_t/2 + \vek l,0\right)\,
S_A\left({\vek k}_t/2 -{\vek l}, \Dr\right).
\end{eqnarray}
The momentum transfer $\vek k_t$ is now shared between 
the two nucleons participating in the interaction. 
This means that  the form factors in Eq.(\ref{eq:coh_ds}) 
are probed, on average, at lower momenta. 
This argument can be applied also to higher order multiple scattering 
amplitudes ($n > 1$) where the momentum transfer is  shared between even 
more nucleons. 
Therefore,  coherent $\rho$ production 
from nuclei is most sensitive to multiple scattering processes when 
$|t|$ is large, a well known feature.

The scattering amplitude 
(\ref{E-FToGamma}) has to be corrected for center-of-mass motion. 
This modification is important especially  for light nuclei. 
For harmonic oscillator wave functions the c.m. correction can be calculated 
exactly \cite{recoil}; 
it leads to an additional factor multiplying the coherent production 
amplitude (\ref{E-FToGamma}): 
\begin{equation}
f_{\gamma^* A \rightarrow \rho A}({\vek k}_t) 
\rightarrow \,
R_{recoil}(\vek k_t^2,\Delta)
\,
f_{\gamma^* A \rightarrow \rho A}({\vek k}_t),  
\end{equation}
with
\begin{equation}
R_{recoil}(\vek k_t^2,\Delta) = 
\exp {\left[\frac{(\vek k_{t}^2+\Delta_{\rho}^2)\,
\langle r^2 \rangle_A}{6 A}\right]}.  
\end{equation}
In numerical calculations we have evaluated the multiple 
scattering series (\ref{eq:coh_exp}) up to $n=2$. 
The corresponding expressions are given in the Appendix.

\subsection{Incoherent production}
\label{ssec:incoh}

The incoherent $\rho$ production cross section is obtained by taking the  difference 
of the inclusive and coherent cross sections: 
\begin{equation}
\frac{d \sigma_{incoh}^A}{d t} = 
\frac{d \sigma_{\gamma^* A \rightarrow \rho X}}{d t} - 
\frac{d \sigma_{\gamma^* A \rightarrow \rho A}}{d t} .  
\end{equation}
Multiple scattering contributions are defined via 
Eqs.(\ref{eq:up_to_n},\ref{eq:ms_coh}):
\begin{equation} \label{eq:incoh_mult}
\frac{d \sigma_{incoh}^{A (n)}}{d t} = 
\frac{d \sigma_{\gamma^* A \rightarrow \rho X}^{(n)}} {d t} - 
\frac{d \sigma_{\gamma^* A \rightarrow \rho A}^{(n)}}{d t}.  
\end{equation}
In the absence of re-scattering, i.e. for single scattering, 
one finds from Eqs.(\ref{eq:incl_sing},\ref{eq:coh_sing}): 
\begin{equation} \label{eq;inc_single}
\frac{d \sigma_{incoh}^{A (0)}}{d t} 
= 
A\,\frac{d \sigma_{\gamma^* N \rightarrow \rho N}}{d t} 
\,\left[1 - S_A({\vek k}_t, \Dr)^2 + 
A S_A({\vek k}_t, \Dr)^2 (1- R_{recoil}^2(\vek k_t^2,\Delta)) 
\right].  
\end{equation}
We observe that, even in the absence of re-scattering,  
the incoherent $\rho$ production cross section from a nucleus 
does not reduce to the production cross section from a single nucleon 
times the number of nucleons inside the target nucleus.
The interference of production processes occuring on different nucleons 
(see discussion in Section~\ref{ssec:incl}),  
as well as nuclear recoil,  leads to a reduction of the nuclear
cross section. 
Nevertheless, for heavy nuclei and/or large momentum transfers 
the nuclear form factors in Eq.(\ref{eq;inc_single}) are small 
and one obtains approximately    
${d \sigma_{incoh}^{A (0)}}/{d t} 
\approx
A\,{d \sigma_{\gamma^* N \rightarrow \rho N}}/{d t}$.

In our numerical studies we calculate the multiple scattering contributions 
up to $n=2$. 
A discussion of their properties and a comparison with 
data from HERMES follows in Section \ref{sec:Results}.

\subsection{Vector meson dominance}

The recent HERMES data on incoherent $\rho$ production from nuclei  
were taken in the kinematic range $0.4\,\mbox{GeV}^2 < Q^2 < 5 \,\mbox{GeV}^2$
and $9 \,\mbox{GeV} < \nu < 20 \,\mbox{GeV}$. 
At these moderate photon energies the observed nuclear effects 
should be caused primarily  by the coherent multiple scattering  of the $\rho$
meson with several nucleons 
(see the discussion in Section \ref{sec:amplitudes_cross sections}). 
It is therefore justified to model the $\rho$-nucleon 
scattering amplitude, $f_{\rho}$, using  Vector Meson Dominance (VMD). 

We  also employ VMD to describe the (virtual) photoproduction 
of the $\rho$ meson from individual nucleons. 
Assuming VMD to be valid, the $\rho$ production amplitude
is related to the scattering amplitude by a kinematic prefactor.
When discussing nuclear phenomena in terms of transparency ratios,
i.e. the ratios of nuclear and free nucleon production
cross sections, details of the VMD photoproduction amplitude
cancel, in particular those parts which would give an inadequate
description at $Q^2 > 1$ GeV$^2$.

VMD relates  the virtual photoproduction amplitude 
$f_{\gamma^*\rho}$ and the elastic $\rho$-nucleon scattering amplitude 
$f_{\rho}$ \cite{BSYP}. For transversely polarized photons one has:  
\begin{equation} \label{eq:fgrT_vmd}
f_{\gamma^*_T\rho} = \frac{\sqrt{\alpha_{em} \pi}}{g_{\rho}}\frac{m_{\rho}^2}
{m_{\rho}^2 + Q^2} f_{\rho},
\end{equation}
with the electromagnetic coupling constant $\alpha_{em}=1/137$, the 
$\rho$ meson mass  $m_{\rho} = 0.77$ GeV, and the coupling constant 
$g_{\rho} = 5.0$. 
The production amplitude for longitudinally polarized photons is given by:
\begin{equation} \label{eq:fgrL_vmd}
f_{\gamma^*_L \rho} = \xi {\sqrt{Q^2}\over m_\rho}
f_{\gamma^*_T \rho}, 
\end{equation}
where $\xi \approx 0.7$  
is the ratio of the elastic longitudinal and transverse 
vector meson-nucleon scattering amplitudes (see e.g. \cite{FSC}). 
It remains to fix the amplitude $f_{\rho}$. 
We use  the parametrization: 
\begin{equation}
\label{E-f_rho_nucleon}
f_{\rho}({\vek k_t}) = 
\frac{|\vek q|}{4 \pi}\sigma_{\rho N} \, (i+\beta) 
e^{-B  \vek k_t^2},
\end{equation}
with the total $\rho$-nucleon scattering cross section 
$\sigma_{\rho N}$, the ratio of the real to imaginary parts  
of the scattering amplitude, $\beta$, and the slope parameter 
$B$. 

For the photoproduction amplitudes  (\ref{eq:fgrT_vmd},\ref{eq:fgrL_vmd})
we use empirical parameters which will be discussed in the following section.

%For the photoproduction amplitudes (\ref{eq:fgrT_vmd},\ref{eq:fgrL_vmd})  
%we use empirical parameters. 
%In the kinematic range of the HERMES experiment we take: 
%$\sigma_{\rho N} = 25$ mb \cite{BSYP} and 
%$\beta  = -0.2$. For the slope parameter we use  
%$B(Q^2 = 1 \,\rm{GeV}^2) \approx 8$ GeV$^{-2}$ and 
%$B(Q^2 = 5 \,\rm{GeV}^2) \approx  6$ GeV$^{-2}$ 
%\cite{BSYP,Ackerstaff:1999wt,Crittenden:1997yz}. 
%We interpolate between these values with the ansatz  
%$B(Q^2) = \frac{1}{3}\left(r_N^2 + r_{q\bar q}^2(Q^2)\right)$ 
%which relates $B$ to the radius of the target nucleon 
%and the $Q^2$-dependent size of the $q\bar q$ Fock state  of the interacting 
%photon. 
%We choose $r_N = 0.8$ fm and $r^2_{q\bar q} = a/(b + Q^2)$ with  
%constants $a = 89.9$ and $b=5.9$ GeV$^2$.

%The vector meson scattering amplitude (\ref{E-f_rho_nucleon}) 
%also describes re-scattering processes. 
%Here a constant slope parameter $B_{\rho} = 8$ GeV$^{-2}$ is 
%appropriate in the kinematic range of the HERMES experiment.  

Since the ratio of longitudinally to transversely polarized photons 
is determined by the kinematics of the process (see e.g. \cite{BSYP}) 
we have now everything 
prepared to calculate nuclear production cross sections. 

While it is not the aim to calculate polarized cross sections in this work,
we do note that polarization is an important observable against which models
can be tested (see e.g. \cite{Deuteron1, Deuteron2}).

\subsection{Discussion}

Before presenting results we discuss 
points at which our present 
calculation differs from previous calculations.
The standard treatment of vector meson electroproduction  
from nuclei within the framework of VMD and Glauber 
multiple scattering theory is summarized in 
Refs.\cite{BSYP,Yennie71,Kop,Hufner:1996dr}. 
Several simplifying but not always justified approximations are 
commonly made. 
It is usually assumed that all momentum is transfered to the target 
nucleus during the production of the vector meson from  one of the nucleons.
After that the propagation of the vector meson through the nuclear 
medium is taken in forward direction.
However, in a treatment of the multiple scattering series 
in Eq.(\ref{E-Nprofile}) without further simplifying assumptions, 
as carried out here, one naturally encounters contributions 
to the production process with the momentum transfer {\it shared} 
between all participating nucleons. 
These contributions, which are usually neglected,  
become increasingly important with rising  $|t|$ .

A further approximation common to most treatments of coherent vector meson 
production 
is the exponentiation of the multiple scattering series. 
While this is a reasonable approximation for heavy nuclei, it should not be 
used for light ones.

In the incoherent case it is usually assumed that the vector meson is produced 
on one specific nucleon. 
Interferences of production processes which occur 
on different nucleons are omitted, 
so that contributions 
proportional to nuclear form factors 
(see  e.g. (\ref{eq:incl_sing})) 
are absent.
Exponentiation of the multiple scattering series is frequently done 
also for the case of incoherent vector meson production. 
This approximation ignores contributions in which 
the momentum transfer is shared between several interacting nucleons.

So far, only total production cross 
sections have been measured with admittedly large experimental 
uncertainties \cite{Ackerstaff:1999wt,Adams:1995bw} so that the  
simplifying approximations,  used 
e.g. in Ref.\cite{Hufner:1996dr},  are not very harmful.
However, as soon as one wants to  describe differential cross sections 
in a more quantitative way a complete term-by-term treatment of multiple 
scattering processes, as done here, is necessary since the $|t|$-dependence
of the cross section is greatly influenced by re-scattering processes.

\section{Results}
\label{sec:Results} 

In the actual calculation we have to fix the 
free parameters of the $\rho$-nucleon scattering amplitude, $f_{\rho}$ 
Eq.~(\ref{E-f_rho_nucleon}).
For the total cross section we use $\sigma_{\rho N}$ = 25 mb in the
kinematic range of the HERMES experiment \cite{BSYP}.  We have varied $\sigma_{\rho N}$
in the range between 22 mb and 30 mb, however these variations do not 
lead to dramatic changes in the
results when considering ratios.
The ratio of real to imaginary part of the amplitude was fixed at
$\xi = -0.2$.

It remains to fix the slope parameter $B$ both for production and re-scattering.
In the case of production, several measurements of the diffractive slope
can be found in the literature (\cite{E665, NMC,H1, H1a, ZEUS, ZEUS2,CHIO}).
A general tendency of decreasing slope with increasing $Q^2$ is observed,
however the data are not precise enough to extract the slope
unambiguously. We use the ansatz
$B(Q^2) = \frac{1}{3}\left(r_N^2 + r_{q\bar q}^2(Q^2)\right)$
which relates $B$ to the radius of the target nucleon 
and the $Q^2$-dependent size of the $q\bar q$ Fock state  of the interacting 
photon, 
where $r^2_{q\bar q} = a/(b + Q^2)$ meets the predictions of
perturbative QCD for large $Q^2$. We require this to agree with the
HERMES result within the kinematic range of the experiment and use
$B(Q^2 = 1 \,\rm{GeV}^2) \approx 8$ GeV$^{-2}$ and 
$B(Q^2 = 5 \,\rm{GeV}^2) \approx  6$ GeV$^{-2}$. A fit fixes the
constants $a = 89.9$ and $b=5.9$ GeV$^2$.

For the diffractive slope of the re-scattering process we use a constant
slope parameter $B_{\rho}=$ 8 GeV$^{-2}$ which is appropriate for the
HERMES kinematics and is choosen in such a way that the incoherent
slope, after taking re-scattering modifications into account,
agrees best with the data.

\begin{figure}[t]
\bigskip
\begin{center} 
\epsfig{file=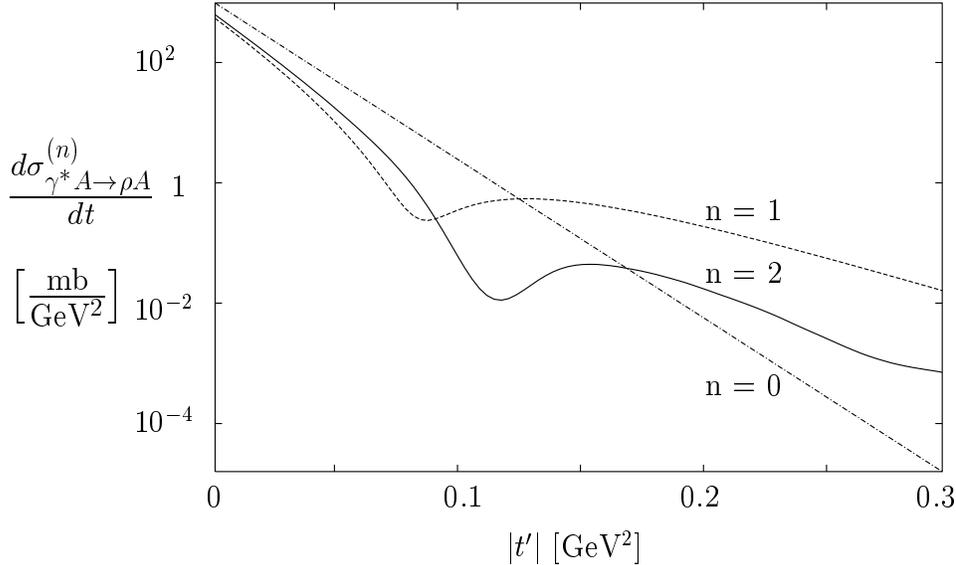,height=80mm}
\end{center}
\caption[...]{Differential cross section  
for coherent $\rho$ production, $\gamma^* A \rightarrow \rho A$, on $^{14}$N at
$\nu = 14$ GeV and $Q^2 = 0.5$ GeV$^2$, plotted against $|t'| = |t - t_0|$ (see text). 
Multiple scattering contributions with up to $n=2$ are shown stepwise.}
\label{fig:coh_cross14N}
\bigskip
\end{figure}

\begin{figure}[t]
\bigskip
\begin{center} 
\epsfig{file=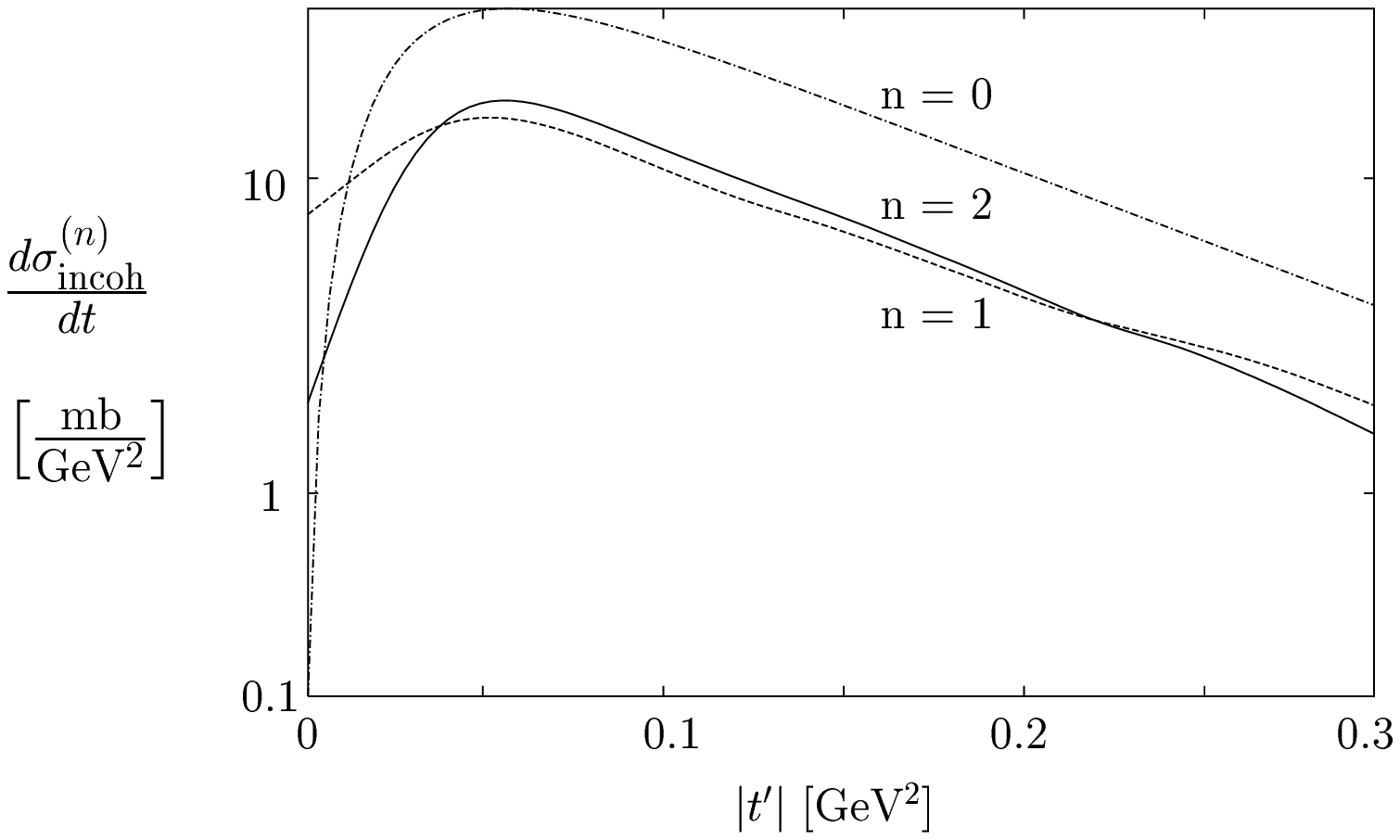,height=80mm}
\end{center}
\caption[...]{Differential cross section for incoherent $\rho$ production, 
$\gamma^* A \rightarrow \rho X$, on $^{14}$N at $\nu = 14$ GeV and $Q^2 = 0.5$
GeV$^2$, plotted versus $|t'| = |t - t_0|$. Multiple scattering contributions 
with up to $n=2$ are shown separately.}
\label{fig:incoh_cross14N}
\bigskip
\end{figure}

Before confronting our calculations with 
data we summarize general features of 
vector meson production cross sections from nuclei. 
Typical results for coherent and incoherent $\rho$ production from   
$^{14}$N are shown in Figs.\ref{fig:coh_cross14N} and \ref{fig:incoh_cross14N}.
The energy and virtuality  of the interacting  photon has been fixed at 
$\nu = 14$ GeV and $Q^2 = 0.5$ GeV$^2$, 
a kinematic range where  VMD  is applicable.

%%%%%%%%%%%%%%%%%%%%%%%%%%%%%%%%%%%%%%%%%%%%%%%%%%%%%%%%%%%%%%%%%%%%
%%%%%%%%%%%%%%%%%%%%%%%%%%%%%%%%%%%%%%%%%%%%%%%%%%%%%%%%%%%%%%%%%%%%%%

In Fig.\ref{fig:coh_cross14N}  we show the multiple scattering 
cross sections $d\sigma^{(n)}_{\gamma^* A \rightarrow \rho A}/dt$ from 
Eq.(\ref{eq:ms_coh})  for coherent $\rho$ production plotted against 
$t' = t - t_0$, with the threshold momentum transfer 
$t_0 \simeq -[(Q^2 + m_\rho^2)/2\nu]^2$. 
We observe that the production cross section is strongly peaked at 
small  $|t| < 0.05$ GeV$^2$ where single scattering dominates.
This behavior is controlled by the nuclear form factor 
present in the single scattering amplitude (\ref{eq:coh_sing}). 
For Gaussian densities one finds 
$d\sigma^{(0)}/dt \sim S_A^2 \sim e^{-{B_A} |t|}$ 
with $B_A = \langle r^2 \rangle_A/3$. 
In the case of  $^{14}$N one has $\langle r^2 \rangle_A^{1/2} = 2.6$ 
fm \cite{Jaeger}, leading 
to $B_A \simeq 57$ GeV$^{-2}$.

With increasing $|t|$ multiple scattering gains in  importance.
At $0.05 \,\mbox{GeV}^2 < |t| < 0.15  \,\mbox{GeV}^2$ the interference between 
single and double scattering leads to a significant reduction of the production
cross section. 
For larger values of $|t|$ multiple scattering adds substantially to
the single scattering cross section and eventually dominates.

Incoherent $\rho$ production cross sections  $d\sigma^{A (n)}_{incoh}/dt$ 
from Eq.(\ref{eq:incoh_mult}) are shown in Fig.\ref{fig:incoh_cross14N}. 
One finds a much weaker dependence on the momentum transfer $|t|$ as compared 
to the coherent case.  
An investigation of single and multiple scattering contributions 
from Section~\ref{ssec:incl}  shows that the $t$-dependence 
is governed here by the behavior of the  virtual photon-nucleon 
and $\rho$-nucleon scattering cross sections. 
These are, in a first approximation, proportional to $e^{-B|t|}$ with 
$B \simeq 7$ GeV$^{-2}$. 
Finally, a comparison of single and multiple scattering cross sections  
shows that single scattering dominates the incoherent production 
cross section for $|t| < 0.5$ GeV$^2$. In this region multiple 
scattering reduces the cross section.

%%%%%%%%%%%%%%%%%%%%%%%%%%%%%%%%%%%%%%%%%%%%%%%%%%%%%%%%%%%%%%%%%%%%
%%%%%%%%%%%%%%%%%%%%%%%%%%%%%%%%%%%%%%%%%%%%%%%%%%%%%%%%%%%%%%%%%%%%%%

In summary: coherent vector meson production from nuclei 
at  $|t| > 1/\langle r^2 \rangle_A$ provides  
an optimal window for an investigation of coherence phenomena.
In this region the cross section is most sensitive 
to multiple scattering. 
However, the coherent production cross 
section is already quite small here -- several orders of magnitude 
smaller than the incoherent one at large $|t|$.

\subsection{Exclusive $\rho$ production at HERMES} 
\label{ssec:HERMES}

The HERMES collaboration has recently produced data \cite{Ackerstaff:1999wt} 
on exclusive electroproduction of 
$\rho$ mesons from protons, deuterium, $^{3}$He, and $^{14}$N. 
The range of energy and momentum transfer available in this experiment is 
$9\,\mbox{GeV} < \nu < 20 \,\mbox{GeV}$ and 
$0.4 \,\mbox{GeV}^2 < Q^2 < 5 \,\mbox{GeV}^2$. 
This implies coherence lengths for $\rho$ mesons, 
\begin{equation}\label{coh_rho}
\lambda_{\rho} = \frac{2\nu}{m_\rho^2 + Q^2}\,,  
\end{equation}
in the range 
$0.6\,\mbox{fm} \,< \lambda < 8\,\mbox{fm}$
covering scales from the size of individual nucleons up to and beyond nuclear dimensions.

\subsubsection{Cross sections} 

%%%%%%%%%%%%%%%%%%%%%%%%%%%%%%%%%%%%%%%%%%%%%%%%%%%%%%%%%%%%%%%%%%%%
\begin{figure}[t]
\bigskip
\begin{center} 
\epsfig{file=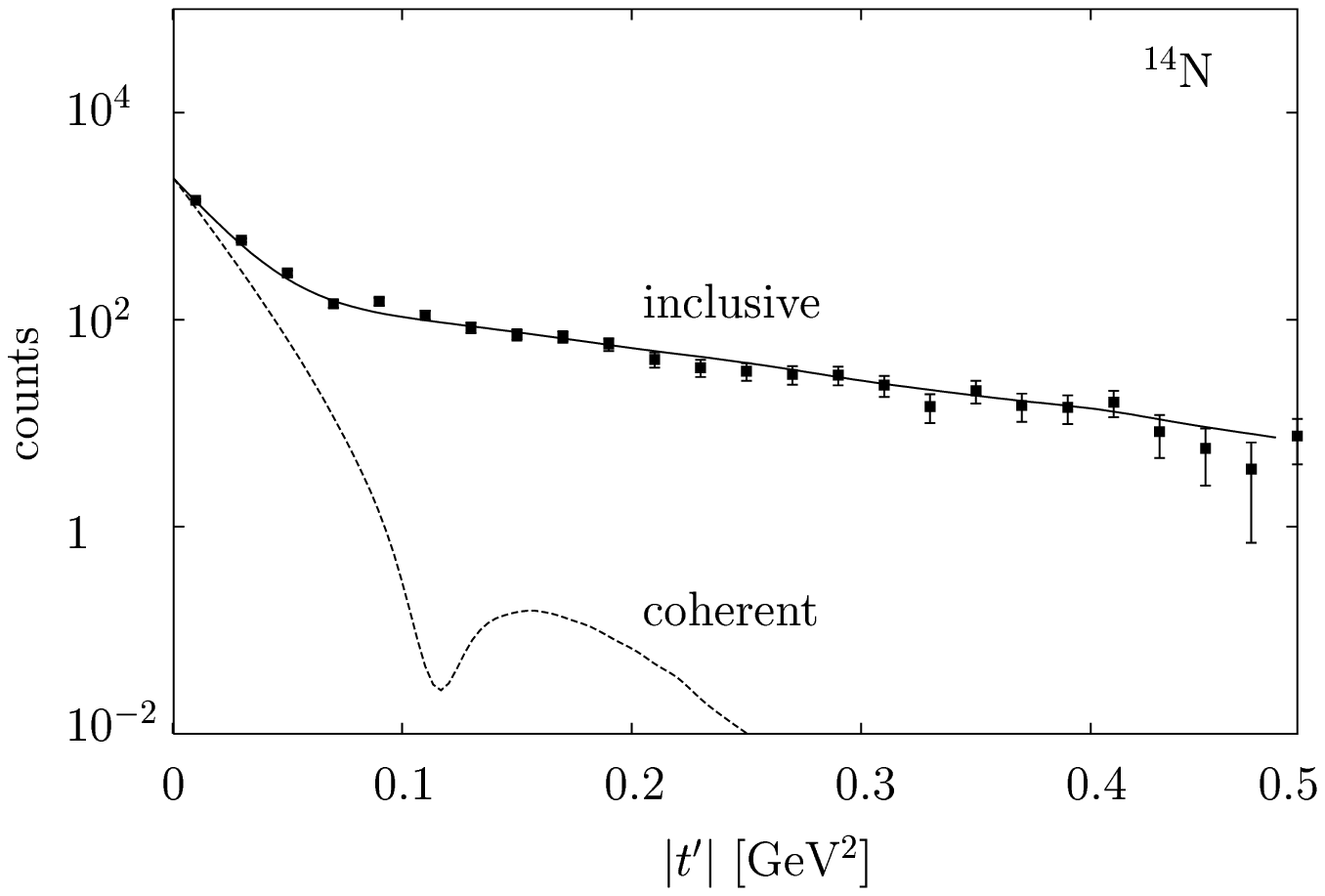,height=60mm}
\epsfig{file=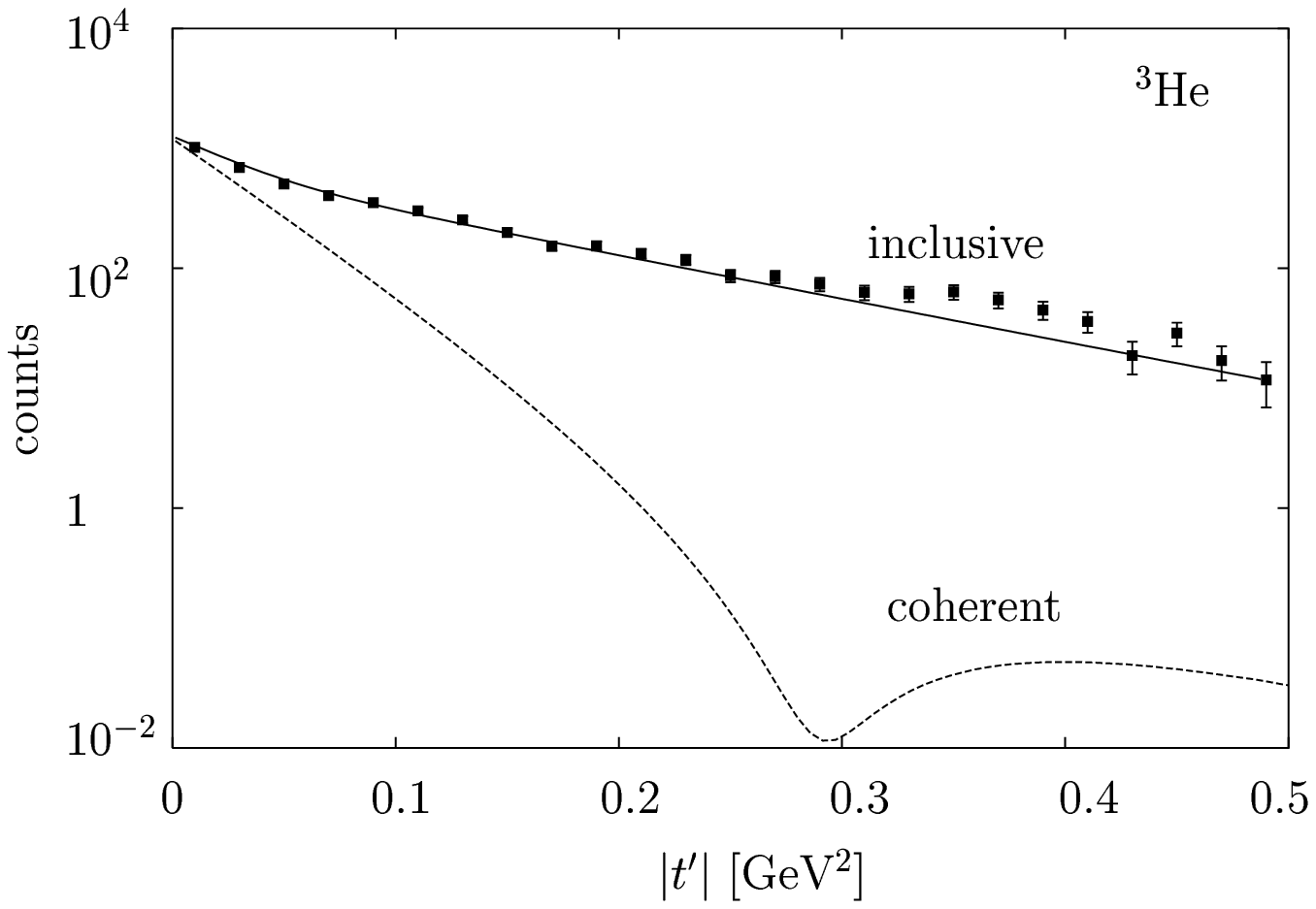,height=60mm}
\end{center}
\caption[...]{Diffractive $\rho$ production from $^{14}$N and 
$^{3}$He. The data are from the HERMES collaboration \cite{Ackerstaff:1999wt}. 
The full (dashed) curves show our results for inclusive (coherent) 
$\rho$ production.
}
\label{fig:rates}
\bigskip
\end{figure}
%%%%%%%%%%%%%%%%%%%%%%%%%%%%%%%%%%%%%%%%%%%%%%%%%%%%%%%%%%%%%%%%%%%%%%

In Fig.\ref{fig:rates}  the rate of produced $\rho$'s is plotted 
for  $^{3}$He  and $^{14}$N against $t$.
At $|t'| =|t-t_0|\ll 0.1$ GeV$^2$ coherent production dominates, leaving the nucleus as
a whole in the ground state. Such coherent processes fall off rapidly 
with the nuclear form factor, so that at  $|t'|\,\gsim\,0.1$ 
GeV$^2$ mostly incoherent $\rho$ production from individual nucleons 
remains.

The result of our calculation,  as 
discussed in Section~\ref{ssec:incl}, is shown for comparison.  
We include contributions to the inclusive production cross section 
(\ref{E-InclusiveAnsatz}) 
up to $n=2$, accounting for the re-scattering from up to $4$ nucleons.
Since no explicit normalization of the data is given in Ref.\cite{Ackerstaff:1999wt}, 
the normalizations of the calculated $^{3}$He  and $^{14}$N cross sections 
have been adjusted to the data in the lowest $|t|$-bin.  
The r.m.s. radii used in these calculations are 
$\langle r^2 \rangle_{^{14}\rm{N}}^{1/2} = 2.6$ fm and 
$\langle r^2 \rangle_{^{3}\rm{He}}^{1/2} = 1.9$ fm \cite{Jaeger}. 
Good agreement with the observed $t$-dependence is found. 

We also show our result for coherent $\rho$ production.  
As expected, at small $|t|$ coherent production 
dominates the inclusive cross section, whereas it drops very rapidly with
rising $|t|$.

\subsubsection{Transparency}

The HERMES collaboration has focused on nuclear effects in incoherent $\rho$ 
production.  For this purpose the transparency ratio 
\begin{equation} \label{eq:TA}
T_A = 
\frac{\sigma_{incoh}^A}{A \, \sigma_{\gamma^* N \rightarrow \rho N}},
\end{equation}
has been analyzed as a function of the  $\rho$ meson propagation length
$\lambda_{\rho}$ from Eq.(\ref{coh_rho}). 
The total incoherent $\rho$ production cross section in 
Eq.(\ref{eq:TA}) is denoted by 
$\sigma_{incoh}^A = \linebreak \int_{-\infty}^0 dt \,d\sigma_{incoh}^A/dt$, 
and  $\sigma_{\gamma^* N \rightarrow \rho N}$ is the total diffractive 
$\rho$ production cross section from free nucleons.
In the absence of re-scattering processes one obtains from 
Eqs.(\ref{eq:incl_sing},\ref{eq:coh_sing}):
\begin{equation} \label{eq:incoh_single}
T_A^{(0)}
\!=\! 
1\! - \! \frac{1}{\sigma_{\gamma^* N \rightarrow \rho N}}
\int_{-\infty}^{0} \!dt  
\frac{d \sigma_{\gamma^* N \rightarrow \rho N}}{d t} \,
S_A({\vek k}_t, \Dr)^2 \!\left[1 - A (1- R^2_{recoil}(\vek k_t^2, \Dr))\right].  
\end{equation}
Incoherent $\rho$ production from $A$ nucleons gives $T_A^{(0)} = 1$. 
A reduction of $T_A^{(0)}$ is caused by the second term in
Eq.(\ref{eq:incoh_single}) which results 
from nuclear recoil and the interference of production amplitudes involving 
different nucleons.

Starting out from the measured inclusive events (Fig.~\ref{fig:rates}) 
the HERMES collaboration has constructed a  data sample 
which approximately represents incoherent 
$\rho$ production according to the following procedure: 
coherent $\rho$ production  processes have been approximated by an exponential
form ($\sim e^{-|t|\,B}$) fitted to the inclusive data at small $|t|$. 
This contribution has then been subtracted from the inclusive events. 
The result so obtained has been  used as  incoherent data sample for 
momentum transfers $|t'_{\lambda_\rho}|< |t'|  < 0.4$ GeV$^2$. 
The lower limit $t'_{\lambda_\rho}$ varies for different bins of the 
$\rho$ meson propagation length. 
It is chosen such that the exponential fit to the small 
$|t|$ data amounts to less than $5\%$ of all inclusive events.

\begin{figure}[t]
\bigskip
\begin{center} 
\epsfig{file=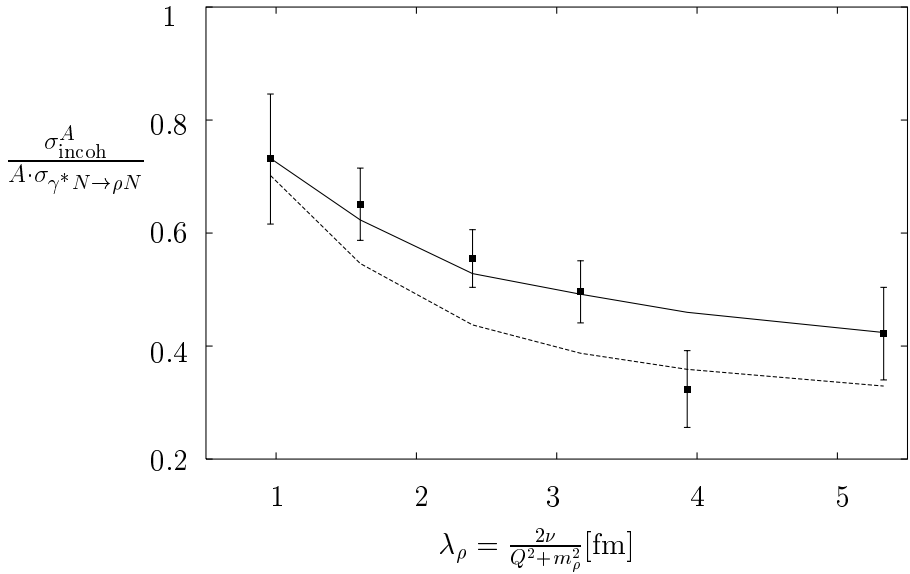,height=80mm}
\end{center}
\caption[...]{The transparency ratio $T_A$ from Eq.(\ref{eq:TA}) for 
$^{14}$N. The data are from the HERMES collaboration \cite{Ackerstaff:1999wt}. 
The full curve shows the result of our calculation which has been 
adapted to the HERMES definition of incoherent events.
The dashed curve shows the transparency ratio as obtained 
from the difference of the calculated inclusive and coherent cross 
section. 
%The single scattering result,  $T_A^{(0)}$ is  indicated by 
%the dotted line.
}
\label{fig:trans}
\bigskip
\end{figure}

The measured transparency ratio $T_{A}$ for $^{14}$N is plotted in
Fig.\ref{fig:trans}. 
The deviation from unity at $\lambda_\rho \, \lsim \, 1$ fm 
comes mainly from the 
final state re-scattering of the $\rho$ meson after being produced on
one of the nucleons.  
With increasing propagation length $\lambda_\rho$ the transparency ratio decreases systematically. 
Hadronic fluctuations of the photon can scatter coherently 
on several nucleons also prior to the production of the final state 
vector meson, leading to an enhancement  of nuclear effects.

%%%%%%%%%%%%%%%%%%%%%%%%%%%%%%%%%%%%%%%%%%%%%%%%%%%%%%%%%%%%%%%%%%%%
%%%%%%%%%%%%%%%%%%%%%%%%%%%%%%%%%%%%%%%%%%%%%%%%%%%%%%%%%%%%%%%%%%%%%%

In Fig.\ref{fig:trans} we present  results of our calculation
within the framework of VMD, using HERMES values for the average 
$Q^2$ and $\nu$ for each data bin. 
The full curve shows our result, adapting the HERMES
construction of incoherent events: 
starting out from the calculated inclusive cross 
section we subtract  an exponential fit to the low-$|t|$ result, and 
use  the HERMES thresholds $t'_{\lambda_\rho}$ as described previously.  
We find good agreement with the data. 
This  suggests that the propagation of initially produced quark-gluon or hadronic states 
through the nuclear environment is dominated by contributions which 
interact about as strongly as the produced 
$\rho$ meson. 
For comparison we also present results for the incoherent transparency ratio as
obtained from the difference of the calculated inclusive and coherent 
cross sections. 
The deviation from the previous result
points at ambiguities in the HERMES definition
of incoherent events.  
%Finally, we show  nuclear effects which present already 
%for single scattering (\ref{eq:incoh_single}). 
%In the case of $^{14}$N and heavier nuclei these turn out to be small. 
%However the situation is different for light nuclei. Here deviations 
%of $T_A^{(0)}$ from unity become important \cite{..}.

Let us have a closer look at the kinematics involved in the HERMES experiment. 
Large propagation length, $\lambda_\rho >  2$ fm, corresponds to data taken at 
high energy transfers $\nu$ accompanied by small $Q^2$. This is the VMD
situation in which the photon predominantly converts into a $\rho$ meson 
which then experiences coherent multiple scattering. 
For small propagation lengths, $\lambda_\rho <  2$ fm, data were taken at 
average values $\bar Q^2$ between $2.5$ and $5$ GeV$^2$ and $\bar \nu \simeq
12$ GeV. At such large $Q^2$, quark-gluon degrees of freedom should already 
be relevant in the initial interaction of the virtual photon with one 
of the nucleons in the target nucleus. 
Perturbative QCD calculations \cite{Frankfurt:1996jw} show indeed that, 
for longitudinally 
polarized photons with $Q^2 \simeq 5$ GeV$^2$, the transverse size 
of the initially produced quark-antiquark wave packet is only a small fraction 
of the diameter of a fully developed $\rho$ meson. 
Still, the time $\tau_f$ to form a $\rho$ meson out of the primary
$q\bar q$ wave packet under these conditions is small.  
Let the invariant mass of the wave packet be of order $\bar Q$. Using 
Eqs.(\ref{eq:ft},\ref{eq:Dhr}) we estimate $\tau_f\sim 2\bar \nu/(\bar Q^2 -
m_\rho^2)< 2.5$ fm/c for $\bar Q > 2.5$ GeV$^2$ and $\bar \nu \simeq 12$ GeV. 
Thus the $q\bar q$ pair is quite likely to be a $\rho$ meson by the time it 
has crossed the average distance between two nucleons in $^{14}$N, and  
its multiple scattering is then governed by the large $\rho N$ (rather 
than the small color-dipole) cross section. 
This is consistent with our observation that the VMD description of the 
measured propagation length effects works well over the full kinematic 
range of the HERMES data.

%interesting insights in the production mechanism: 
%For small $\lambda_\rho < 2$ fm data where taken at average values 
%$2.5 \, \mbox{GeV}^2 < \bar Q^2 < 5 \,\mbox{GeV}^2$ and $\bar \nu \simeq 12$ GeV. 
%In this kinematic region quark-gluon degrees of freedom should already be 
%relevant in the initial interaction of the virtual photon with one of the 
%nucleons in the target nucleus. 
%Indeed, perturbative calculations show that for  longitudinally polarized 
%photons 
%and  $Q^2 \simeq 5\,\T{GeV}^2$  the transverse size of the initially produced
%quark-gluon  wave packet  is  a small fraction of the diameter of a fully
%developed  $\rho$ meson \cite{..}. 
%%Our observation, that the  measured nuclear effects can be described 
%within the framework of VMD shows that the time needed to form 
%the measured $\rho$ from the produced quark-gluon configuration 
%cannot be much larger than  the typical distance between two nucleons 
%inside the target nucleus. 
%Indeed the qualitative estimate for the formation time in Eq.(\ref{eq:ft}) gives  
%$\tau_f \simeq 3$ fm and confirms this picture.

%On the other hand for  $\lambda_\rho \,\gsim \,  4$ fm data where taken at 
%%larger average energies  $\bar \nu \,\gsim\, 15$ GeV leading to formation times 
%larger than the average nucleon-nucleon separation. 
%However here 
%$\bar Q^2 \,\lsim\, 1$ GeV$^2$ and we are clearly in the 
%domain where hadronic configurations in the photon wave functions 
%dominate the production process.

\subsection{Aspects of color singlet transparency} 

Nuclear effects in incoherent $\rho$ production as observed by the HERMES
collaboration result from the coherent re-scattering of hadronic components  
present in the photon spectral function.
On the other hand, at $Q^2 \gg 1$ GeV$^2$ small sized quark-gluon wave packets 
are produced in photon-nucleon collisions. 
Their color dipole moment or, equivalently, the 
interaction cross section vanishes in the limit of large $Q^2$.
Then nuclear effects become negligible and color singlet transparency occurs.

The study  of quark-gluon configurations of the interacting virtual  
photons requires 
a large lever-arm in $Q^2$ and $\nu$. In addition one has to identify a  
kinematic window where the vector meson production process is most sensitive 
to coherent multiple scattering from several nucleons in the target nucleus. 
Finally, the influence of different characteristic scales on the production
process have to be separated \cite{Kop,Deuteron1,Deuteron2}.

%%%%%%%%%%%%%%%%%%%%%%%%%%%%%%%%%%%%%%%%%%%%%%%%%%%%%%%%%%%%%%%%%%%%
\begin{figure}[t]
\bigskip
\begin{center} 
\epsfig{file=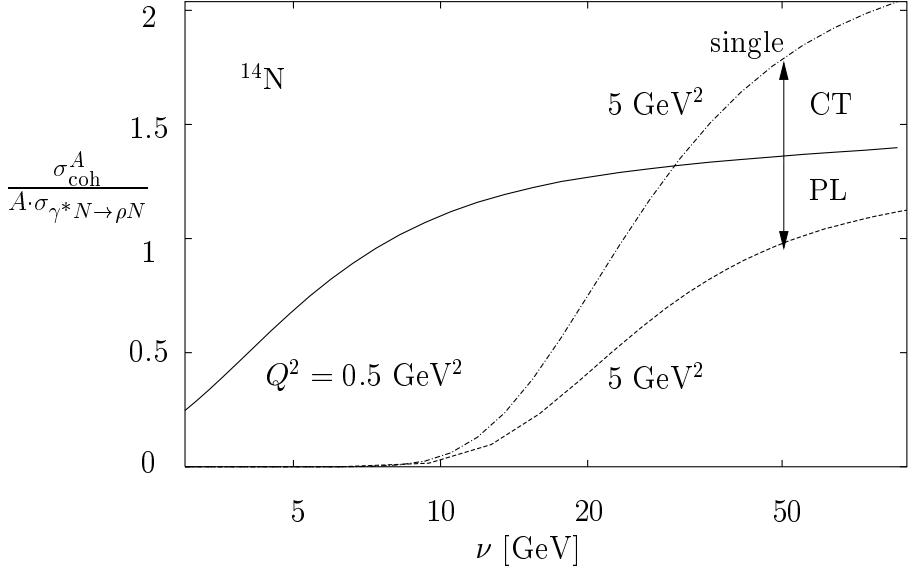,height=80mm}
\end{center}
\caption[...]{The transparency ratio for coherent $\rho$ production 
from $^{14}$N. The full (dashed) curve corresponds to the VMD calculation 
at $Q^2 = 0.5$ GeV$^2$ ($5$ GeV$^2$). The dash-dotted curve shows the single 
scattering result for $Q^2 = 5$ GeV$^2$. The arrows indicate the 
behavior of the transparency ratio for increasing $Q^2$, as caused 
by propagation length effects (PL) and color transparency (CT). 
}
\label{fig:coh_vs_ct}
\bigskip
\end{figure}
%%%%%%%%%%%%%%%%%%%%%%%%%%%%%%%%%%%%%%%%%%%%%%%%%%%%%%%%%%%%%%%%%%%%%%

%%%%%%%%%%%%%%%%%%%%%%%%%%%%%%%%%%%%%%%%%%%%%%%%%%%%%%%%%%%%%%%%%%%%%%
\begin{figure}[t]
\bigskip
\begin{center} 
\epsfig{file=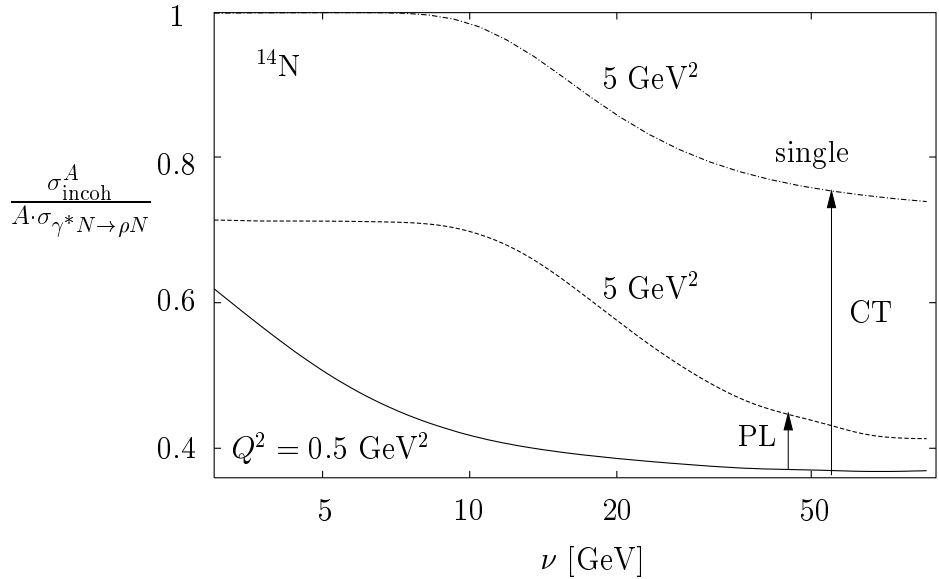,height=80mm}
\end{center}
\caption[...]{The transparency ratio for incoherent $\rho$ production 
from $^{14}$N. The full (dashed) curve corresponds to the VMD calculation 
at $Q^2 = 0.5$ GeV$^2$ ($5$ GeV$^2$). The dash-dotted curve shows the single 
scattering result for $Q^2 = 5$ GeV$^2$. The arrows indicate the 
behavior of the transparency ratio for increasing $Q^2$, as caused 
by propagation length effects (PL) and color transparency (CT). 
}
\label{fig:inco_vs_ct}
\bigskip
\end{figure}
%%%%%%%%%%%%%%%%%%%%%%%%%%%%%%%%%%%%%%%%%%%%%%%%%%%%%%%%%%%%%%%%%%%%%%

For these reasons coherent vector meson production from nuclei turns out to 
be very  useful. 
At $|t| > 1/\langle r^2 \rangle_A$ the corresponding differential cross section
depends crucially on contributions from coherent multiple scattering  
(Figure ~\ref{fig:coh_cross14N}).

Coherent vector meson production is also favorable with respect to a separation
of phenomena caused by variations of the propagation length and color transparency.  
This fact is demonstrated in Fig.~\ref{fig:coh_vs_ct}:  
we compare the transparency ratio 
$\sigma_{\gamma^* A \rightarrow \rho A}/ A \sigma_{\gamma^* N
\rightarrow \rho N}$ for different values of $Q^2$ and $\nu$ as obtained 
within VMD. The decrease of the propagation length with rising $Q^2$ leads apparently 
to a suppression of the cross section ratio. 
If color transparency is fully established at large $Q^2$ only single scattering occurs. 
At large energies $\nu$ this leads to an increase of the transparency ratio
as compared to the VMD scenario. 
An increase of the transparency ratio with rising $Q^2$ is therefore a clear 
signal for the onset of color transparency.

In incoherent vector meson production the situation is different. As
illustrated in Fig.~\ref{fig:inco_vs_ct}, propagation length phenomena and color
transparency lead to a similar trend for the $Q^2$-dependence of the 
transparency ratio. A separation of both effects is therefore difficult.
Similar observations have been made in \cite{Kop}.

The difference between both cases has two reasons. The first one is the reduction
of the cross section by re-scattering contributions, a consequence of
the fact that the scattering amplitudes are dominated by their imaginary parts.
Therefore the color transparency effect which suppresses re-scattering always
leads to a rise in the nuclear cross section relative to that of the
free nucleon. On the other hand any changes in the
propagation length affect primarily the coherent scattering process. Since
$\sigma^A_{\text{incoh}}$ = $\sigma^A_{\text{inclusive}}$ - $\sigma^A_{\text{coh}}$,
the coherent contribution enters with opposite sign when we consider
incoherent scattering. This explains the difference between Fig.~\ref{fig:coh_vs_ct} and
Fig.~\ref{fig:inco_vs_ct}.

With the (simplistic) assumption that color transparency completely suppresses
any re-scattering, its signature should be visible at $\nu=50$ GeV when one systematically 
compares the range in $Q^2$ from 0.5 up to 5 GeV$^2$, at least in our model.
A more detailed calculation requires in addition a model for the
expansion of the initially small color dipole.

\section{Summary}

We have developed the multiple scattering formalism for the electroproduction of
$\rho$ mesons from nuclei. This process is an excellent tool to investigate
the formation, propagation and hadronization of quark-antiquark-gluon
fluctuations of the virtual high-energy photon.

The theory has been compared to the HERMES measurements for the case of
$\rho$ electroproduction on $^{14}$N. The characteristic reduction
of the nuclear transparency with increasing $\rho$ meson propagation
length is well reproduced. The more detailed quantitative analysis
would still require an improved separation of incoherent and
coherent events.

We have also examined the sensitivity with respect to the quest for
color transparency in such processes. While the $A$-dependence
of incoherent production does not permit a clear separation
between color transparency and standard propagation length effects,
coherent $\rho$ electroproduction appears to be well
suited for this purpose.

\section*{Acknowledgements}
We would like to thank L. Frankfurt, M. Strikman, M. Sargsian, G. van
der Steenhoven and T. O'Neill for discussions.

\newpage

\appendix

\section{Amplitudes and cross sections}

\subsection{Coherent amplitudes}

We summarize the amplitudes for coherent $\rho$ electroproduction
used in our calculations up to $n=2$. The simplifying assumption
$\text{Re}f_\rho \ll \text{Im}f_\rho$ has been used.

\begin{equation}
f_{\gamma^{\ast} A \rightarrow \rho A}^{(0)}({\vek k}_t) = A  S_A({\vek k}_t, -\Dr) f_{\gamma^{\ast}\rho}({{\vek k}_t}),
\end{equation}

\begin{equation}
\begin{split}
f_{\gamma^{\ast}A \rightarrow \rho A}^{(1)}({\vek k}_t)=&
i\frac{A(A-1)}{4 \pi |\vek{q}|}
\intn d^2 l \fr{{\vek k}_t/2 + {\vek l}} 
f_{\gamma^{\ast}\rho}({{\vek k}_t/2 - {\vek l}})
\\ & \times 
S_A({\vek k}_t/2 +{\vek l},0) S_A({\vek k}_t/2 - {\vek l}, -\Dr),
\end{split}
\end{equation}

\begin{equation}
\begin{split}
f_{\gamma^{\ast} A \rightarrow \rho A}^{(2)}({\vek k}_t) =&
- \frac{A(A-1)(A-2)}{32 \pi^2 |{\vek q}|^2}
\intn d^2 l_1 \intn d^2 l_2
f_{\rho}({\vek k}_t/3 + {\vek l}_1)
\\ & \times
f_{\rho}({\vek k}_t/3 + {\vek l}_2)
f_{\gamma^{\ast}\rho}({\vek k}_t/3 - {\vek l}_1 -{\vek l}_2)
\\ & \times
S_A({\vek k}_t/3 + {\vek l}_1, 0)
S_A({\vek k}_t/3 + {\vek l}_2, 0)
S_A({\vek k}_t/3 - {\vek l}_1 -{\vek l}_2, -\Dr).
\end{split}
\end{equation}

\subsection{Inclusive cross sections}

Here we give explicit expressions for the incoherent cross 
sections derived from the expansion of 
Eq.(\ref{E-InclusiveAnsatz}) for $n=0,1$. The $n=0$ result reads:

\begin{equation} 
\frac{d \sigma_{\gamma^* A \rightarrow \rho X}^{(0)}}{d t}
= 
A\,\frac{d \sigma_{\gamma^* N \rightarrow \rho N}}{d t} 
\,\left[1 + (A-1) S_A({\vek k}_t, -\Dr) S_A(-{\vek k}_t, \Dr)  \right].  
\end{equation}

The $n=1$ result includes six additional interference terms between production
and re-scattering amplitudes as well as the square of re-scattering amplitudes,

\begin{equation}
\frac{d \sigma_{\gamma^* A \rightarrow \rho X}^{(1)}}{d t} = 
\frac{d \sigma_{\gamma^* A \rightarrow \rho X}^{(0)}}{d t} 
+ \sum_{i = 1}^6 \frac{d \sigma_{\gamma^* A \rightarrow \rho X}^{(1)}}{d t} \Bigg|_{i},
\end{equation}

given by the following expressions:

\begin{equation} 
\begin{split}
\frac{d \sigma_{\gamma^* A \rightarrow \rho X}^{(1)}}{d t}\Bigg|_{1} \,= \, &  
-\frac{A(A-1)}{2 |\vek q|^3} \intn d^2 l
\quad \text{Im} 
\left[ \fgrs{{\vek k}_t} \fgr{{\vek k}_t - {\vek l}} \fr{{\vek l}} \right]
\\ & \times
\left\{
S_A({\vek l},0) S_A({-\vek l},0) + 
S_A({- \vek k}_t+{\vek l}, \Dr) S_A({\vek k}_t - {\vek l}, -\Dr)\right.
\\ 
& 
\;\;\;\;\;
\left.+
(A-2) S_A({\vek k}_t, -\Dr) S_A({- \vek k}_t + l, \Dr) S_A({-\vek l},0) 
\right\},
\end{split}
\end{equation}

\begin{equation}
\begin{split}
\frac{d \sigma^{(1)}_{\gamma^{\ast}A \rightarrow \rho X}}{d t}\Bigg|_{2} &=
\frac{A(A-1)}{8 \pi |{\vek q}|^4}
\intn d^2 l_1 \intn d^2 l_2
f_{\gamma^{\ast}\rho}^{\ast}({\vek k}_t-{\vek l}_1)
f_{\gamma^{\ast}\rho}({\vek k}_t - {\vek l}_2) \qquad \qquad \qquad 
\\ & \times
f_{\rho}^{\ast}({\vek l}_1) \fr{{\vek l}_2}
S_A({\vek l}_1-{\vek l}_2 , 0) S_A({\vek l}_2-{\vek l}_1 , 0),
\end{split}
\end{equation}

\begin{equation}
\begin{split}
\frac{d \sigma^{(1)}_{\gamma^{\ast}A \rightarrow \rho X}}{d t}\Bigg|_{3} &=
\frac{A(A-1)(A-2)}{8 \pi |{\vek q}|^4}
\intn d^2 l_1 \intn d^2 l_2 
f_{\gamma^{\ast}\rho}^{\ast}({\vek k}_t-{\vek l}_1) 
f_{\gamma^{\ast}\rho}({\vek k}_t- {\vek l}_2) \qquad 
\\ & \times
\frs{{\vek l_1}} \fr{{\vek l_2}}
S_A({\vek l}_1-{\vek l}_2,0) S_A(-{\vek l}_1,0) S_A({\vek l}_2,0),
\end{split}
\end{equation}

\begin{equation}
\begin{split}
\frac{d \sigma^{(1)}_{\gamma^{\ast}A \rightarrow \rho X}}{d t}\Bigg|_{4} &=
\frac{A(A-1)(A-2)}{12 \pi |{\vek q}|^4}
\intn d^2 l_1 \intn d^2 l_2 
f_{\gamma^{\ast}\rho}^{\ast}({\vek k}_t -{\vek  l}_1)
f_{\gamma^{\ast}\rho}({\vek k}_t -{\vek l}_2) 
\\ & \times
\frs{{\vek l}_1} \fr{{\vek l}_2}
S_A(-{\vek k}_t +{\vek l}_1,\Dr) S_A({\vek k}_t - {\vek l}_2,-\Dr)
S_A({\vek l}_2-{\vek l}_1), 
\end{split}
\end{equation}

\begin{equation}
\begin{split}
\frac{d \sigma^{(1)}_{\gamma^{\ast}A \rightarrow \rho X}}{d t}\Bigg|_{5}  &=
\frac{A(A-1)(A-2)(A-3)}{16 \pi |{\vek q}|^4}
\\ & \times
\intn d^2 l_1 \intn d^2 l_1
f_{\gamma^{\ast}\rho}^{\ast}({\vek k}_t - {\vek l}_1) 
f_{\gamma^{\ast}\rho}({\vek k}_t - {\vek l}_2)
\frs{{\vek l}_1} \fr{{\vek l}_2} \qquad \qquad \qquad 
\\ & \times
S_A(-{\vek k}_t + {\vek l}_1,\Dr) S_A({\vek k}_t -{\vek l}_2,-\Dr)
S_A(-{\vek l}_1,0) S_A({\vek l}_2 ,0),
\end{split}
\end{equation}

\begin{equation}
\begin{split}
\frac{d \sigma^{(1)}_{\gamma^{\ast}A \rightarrow \rho X}}{d t}\Bigg|_{6}  &=
\frac{A(A-1)(A-2)}{24 \pi |{\vek q}|^4}
\\ & \times
\intn d^2 l_1 \intn d^2 l_2
f_{\gamma^{\ast}\rho}^{\ast}({\vek k}_t - {\vek l}_1)
f_{\gamma^{\ast}\rho}({\vek k}_t - {\vek l}_2) 
\frs{{\vek l}_1} \fr{{\vek l}_2} \qquad \qquad \qquad 
\\ & \times
S_A(-{\vek k}_t +{\vek l}_1 - {\vek l}_2,\Dr) S_A({\vek k}_t -{\vek l}_1,-\Dr)
S_A({\vek l}_2,0).
\end{split}
\end{equation}

The additional contributions to the $n=2$ result, not shown here, 
are also included in the numerical
calculations. They are constructed in exactly the same way as the
expressions already shown by accounting for all possible combinations
of struck nucleons.

%%%%%%%%%%%%%%%%%%%%%%%%%%%%%%%%%%%%%%%%%%%%%%%%%%%%%%%%%%%%%%%%%%

\end{document}